\documentclass[aps,prl,twocolumn,showpacs,preprintnumbers,amsmath,amssymb,floatfix]{revtex4-2}

\usepackage{graphics}
\usepackage{graphicx}
\usepackage{amsmath}
\usepackage{setspace}
\usepackage{braket}
\usepackage{epstopdf}
\usepackage{comment}
\usepackage{hyperref}
\usepackage{bm}
\usepackage{xfrac}
\usepackage{xcolor}

\usepackage{mathrsfs}
\hypersetup{%
   pdfpagemode=None, 
   pdfstartpage=1,
   pdfmenubar=true,
   pdftoolbar=true,
   colorlinks = true,
   linkcolor=blue,
   citecolor=blue,
   urlcolor=blue,
   bookmarksopen=false
 }
\usepackage[capitalize]{cleveref} 
\crefname{figure}{FIG.}{FIGs.}

\begin{document}


\author{Marjan Mirahmadi$^1$, Jes\'us P\'erez-R\'ios$^{1,2}$, Oleg Egorov$^3$, Vladimir Tyuterev$^{3,4}$, Viatcheslav Kokoouline$^5$}
\affiliation{%
$^1$ Fritz-Haber-Institut der Max-Planck-Gesellschaft, Berlin, Germany\\
$^2$Department of Physics, Stony Brook University, Stony Brook, New York 11794, USA\\
$^3$Quamer Laboratory, Tomsk State University, Tomsk, Russia\\
$^4$Groupe de Spectrometrie Mol\'eculaire et Atmospherique, UMR CNRS 7331, University of Reims Champagne-Ardenne, Reims, France\\
$^5$Department of Physics, University of Central Florida, Florida, USA
}%

\title{
Ozone formation in ternary collisions: Theory and experiment reconciled
}

\date{\today}

\begin{abstract}
\hyphenchar\font=-1

Absorbing UV radiation, ozone protects life on Earth and plays a fundamental role in Earth's temperature balance. The formation of ozone occurs through the ternary recombination reaction: O$_2$ + O + M $\rightarrow $O$_3$ + M, where M can be N$_2$, O$_2$ or Ar. Here, we developed a theoretical approach capable of modeling the formation of ozone molecules in ternary collisions, and applied it to the reaction with M=Ar because of extensive experimental data available. The rate coefficients for the direct formation of O$_3$ in ternary collisions O+O$_2$+Ar were computed for the first time as a function of collision energy, and thermally-averaged coefficients were derived for temperatures 5-900~K leading to a good agreement with available experimental data for temperatures 100-900~K. The present study shows that the formation of ozone in ternary collisions O+O$_2$+Ar at temperatures below 200~K proceeds through a formation of a temporary complex ArO$_2$, while at temperatures above 1000~K, the reaction proceeds mainly through a formation of long-lived vibrational resonances of O$_3^*$. At intermediate temperatures 200~K-1000~K, the process cannot be viewed as a two-step mechanism. In addition, it is found that the majority of O$_3$ molecules formed initially are weakly bound.


\end{abstract}

\maketitle


\section{\label{sec:intro}Introduction}
Being the major absorber of the UV light in the upper atmosphere \cite{orphal2016absorption}, the ozone molecule is crucial for the well-being of humanity. However, the tropospheric ozone, the concentration of which near the Earth’s surface was doubled or tripled in the troposphere during XX century \cite{IP01} is harmful to the human respiratory system and, therefore, is considered as an air pollutant. Therefore, from an ecological point of view, ozone should be protected from destruction in the upper atmosphere, but its production in living or workspaces must be minimized. Furthermore, the O$_3$/O$_2$ photochemical cycle has a significant impact on the chemistry in the Earth’s atmosphere and climate change \cite{barnes2019ozone}.


In Earth's ionosphere, formation of ozone occurs through the reaction of ternary recombination: O$_2$ + O + M $\rightarrow $O$_3$ + M, where M is typically N$_2$, O$_2$. Knowledge of the “nascent population”  of ozone is essential for reliable modeling of the satellite measurements in nonlocal thermodynamic conditions of the upper atmosphere \cite{l2001non,Kaufmann2006,feofilov2012infrared}. In many laboratory experiments \cite{hippler1990temperature,rawlins1987dynamics,LUT05:2764,janssen2001kinetic} it was shown that the ternary rate coefficient  $k_3$  for M=N$_2$ have a very similar behavior as those of M=Ar. A complete database is available \cite{LUT05:2764} for the argon colliding partner, which is thus often used to study ozone formation.

In the existing literature, two simplified mechanisms were proposed to analyze the process. Both mechanisms proceed in two steps forming an intermediate complex: stabilization and Chaperon mechanisms. In the stabilization mechanism (also known as energy transfer mechanism), the complex is a long-lived rovibrational resonance O$_3^*$ of ozone, whereas, in the Chaperon mechanism, the complex is a resonance of O$_2$M$^*$. The relative importance of the two mechanisms in ozone formation at particular conditions is determined by (a) lifetimes and densities of resonances in the energy interval corresponding to collision energies and (b) partial pressures of O, O$_2$, and M.  It is important to stress that the rate coefficient for the ternary recombination, properly determined theoretically or, in an ideal world, experimentally, accounts for both (and any other) two-step mechanisms. Therefore, the ternary recombination rate coefficient can be used to analyze the importance of one of the two-step mechanisms at given densities. 

All previous theoretical studies of ternary recombination of ozone have implied one or two two-step mechanisms for ozone formation. The novelty of the present work is that it aims at obtaining the rate coefficient considering the ternary collision O$_2$ + O + M without splitting the process into two steps. The present theoretical approach combines an {\it ab initio} potential energy surfaces (PES) of the ArO$_3$ system, obtained in this study, and the classical-trajectory method in hyperspherical coordinates developed previously by two of us {\it et al.} \cite{Perez-Rios2014,Greene2017,Mirahmadi2021}.

\section{Results}
\begin{figure}
\centering
\includegraphics[width=0.48 \textwidth]{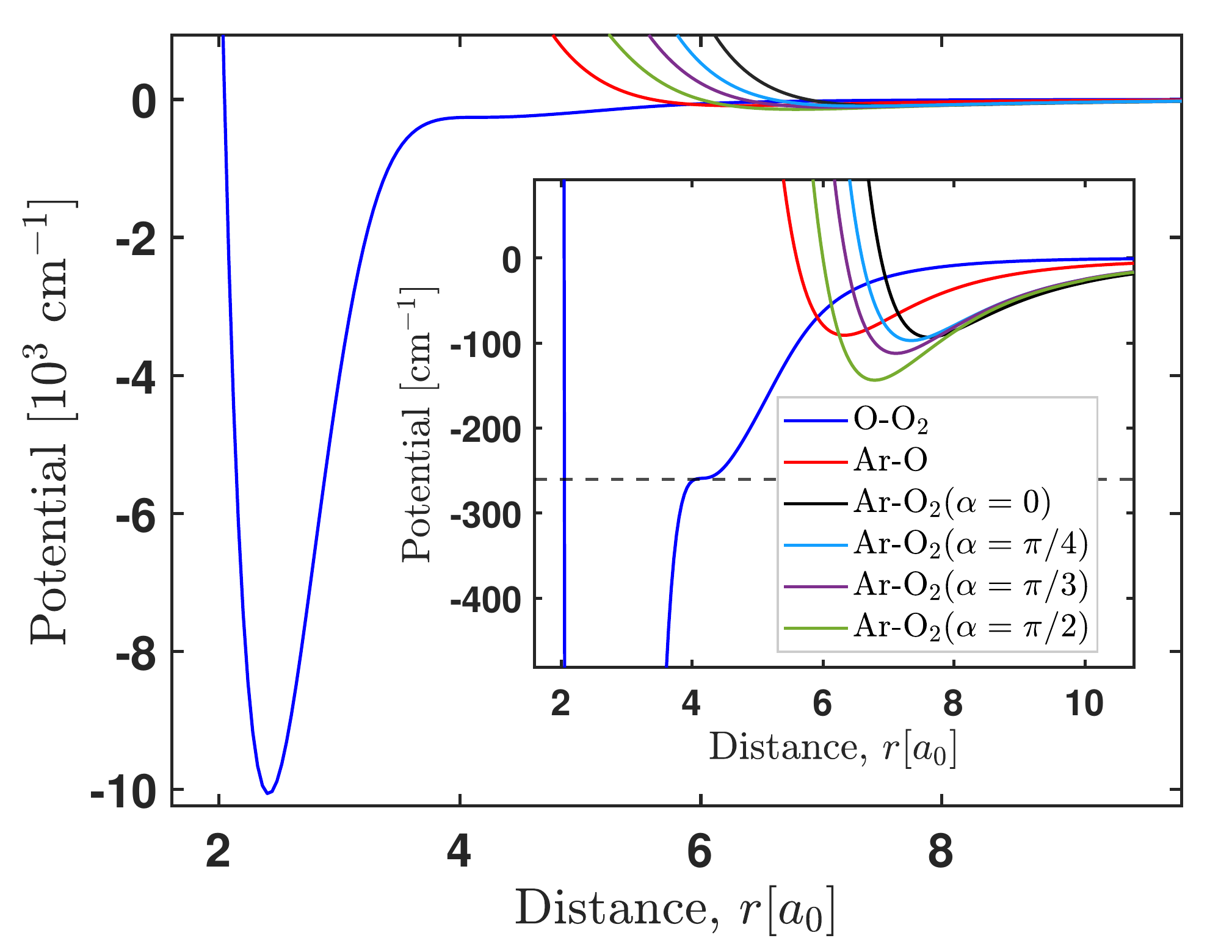}
\caption{Relevant interaction potentials for the reaction O$_2$ + O + Ar $\rightarrow$ O$_3$ +Ar. Dissociation energies are $D_e^{\mathrm{O}-\mathrm{O}_2}= 1.04 \times 10^4$~cm$^{-1}(\approx 1.5 \times 10^4$ K) and $D_e^\mathrm{Ar-O}$=90.7967~cm$^{-1}$ ($\approx 131$~K). For for $\alpha= 0$, $D_e^\mathrm{Ar-O_2}$=93.13~cm$^{-1}$ ($\approx 134$~K) and for $\alpha= \pi/2$,
$D_e^\mathrm{Ar-O_2}$=143.6462~cm$^{-1}$ ($\approx 207$~K). The black dashed line in the inset indicates the energy of the shoulder structure of O-O$_2$ at $\approx 260$ cm$^{-1}$ ($\approx 374$~K).}
\label{fig:pots}
\end{figure}

{ArO$_3$ \it potential.}
The potential energy surface (PES) used in this study is constructed as the sum $V=V_\mathrm{O-O_2}+V_\mathrm{Ar-O_2}+V_\mathrm{Ar-O}$. $V_\mathrm{O-O_2}$  is the intramolecular {\it ab initio} potential energy surface (PES) of ozone \cite{TYU13:134307}.  The high accuracy of the PES was confirmed by the study of ro-vibrational levels near the dissociation threshold \cite{TYU14:143002} and accounting for the complete permutation symmetry of identical nuclei \cite{O3_PCCP,valijch2020}, as well as by quantum calculations of isotopic exchange reactions both in stationary \cite{HON18:214304} and time dependent \cite{YUE19:7733} approaches. $V_\mathrm{Ar-O_2}$ and $V_\mathrm{Ar-O}$ are the intermolecular {\it ab initio} PESs built in this study (see the Methods for details). The constructed PESs have the following configurations of the global minima:  $R_e^\mathrm{Ar-O}$=6.326~$a_0$ (where $a_0 \approx 5.29\times~10^{-11}$m is the Bohr radius) and $R_e^\mathrm{Ar-O_2}$=6.787~$a_0$ ($\alpha= \pi/2$) with the depth of $D_e^\mathrm{Ar-O_2}$=143.6462~cm$^{-1}$ ($\approx 207$~K) and $D_e^\mathrm{Ar-O}$=90.7967~cm$^{-1}$ ($\approx 131$~K), respectively. Potentials are shown in Fig.~\ref{fig:pots}.

{\it Dynamics.}
Rigorous full quantum scattering calculations for the system are considered as practically unfeasible \cite{SCH06:625,TEP16:19194}. Various classical calculations \cite{schinke2005,ivanov2006}, -- statistical  \cite{GAO01:259} or {\it ad hoc} simplified quantum models and essentially focused on one of the two-step mechanisms,-- have been reported \cite{CHA04:2700,XIE05:131,GRE09:181103,TEP18:259}, but a qualitative agreement is still lacking, as reviewed in Refs.~\cite{SCH06:625,TEP16:19194}.

The present approach is based on a classical trajectory (CT) method in hyperspherical coordinates, which has shown efficiency for reactive collisions of three neutral atoms \cite{Perez-Rios2014,Greene2017,Mirahmadi2021,Mirahmadi2021a} as well as for ion-neutral-neutral ternary recombination processes \cite{Perez-Rios2015,Krukow2016,JPR2018}. The method is adapted to the present problem representing the O$_2$ molecule as a super-atom. Thus, reducing the degrees of freedom of the O+O$_2$+Ar system. In calculations, it results in a fixed angle between O$_2$ and the direction to O, corresponding to the bond angle in O$_3$ (117$^\circ$). The interaction of Ar with O$_3$ does not depend significantly on the angle between the plane of O+O$_2$ and the direction to Ar, so to simplify the numerical part of the approach, the angle is fixed to 0. Finally, the angle $\alpha$ between O$_2$ and Ar is considered a parameter of the problem; Calculations are made for several values of $\alpha$ and the result is averaged over that parameter (see the Methods for details). 



The application of this method to the formation of ozone via ternary recombination reaction $\mathrm{O}_2 + \mathrm{O} + \mathrm{Ar} \rightarrow \mathrm{O}_3 + \mathrm{Ar}$ is based on the assumption that the internal degrees of freedom of the oxygen molecule do not play an essential role since the excitation of the O$_2$ vibrational mode requires high collision energy, i.e., over 1000~cm$^{-1}$($\approx$~1439~K). 
The assumption can be verified calculating rate coefficients for different bond angles of the ArO$_2$ molecule, $\alpha$.
As a result, we are able to estimate the uncertainty of the present calculations due to the approximation of the frozen internal degrees of freedom of the oxygen molecule (see the Methods for details).


Finally, the thermal average of the three-body recombination rate coefficient is obtained by integrating the energy-dependent three-body recombination rate coefficient, $k_3(E_c)$, over the appropriate three-body Maxwell-Boltzmann distribution of collision energies, $E_c$, yielding
\begin{equation}\label{eq:MBa}
	k_3(T) =  \frac{1}{2(k_B T)^3} \int_{0}^{\infty}k_3(E_c) E_c^2 e^{-E_c/(k_BT)} dE_c ~,
\end{equation}
where $k_B$ is the Boltzmann constant and $T$ the temperature of the system.


\begin{figure}
	\centering
	\includegraphics[scale=0.42]{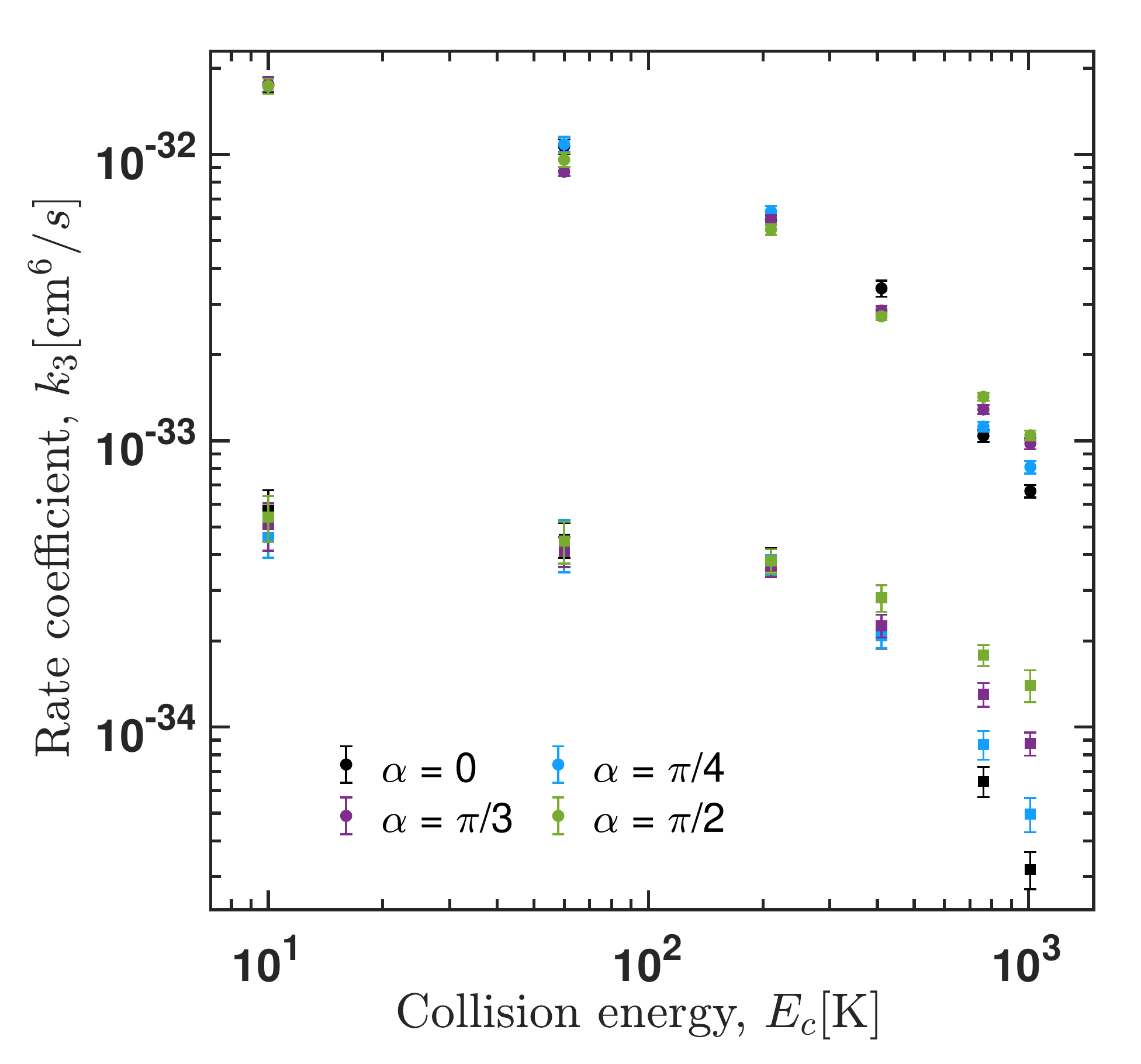}
	\caption{Energy-dependent recombination rates for O$_3$ molecules of all binding energies (circles) and those with binding energies $>200 ~\mathrm{cm}^{-1}$ (squares) for different values of $\alpha$. The energy (abscissa) is given in temperature units.}
	\label{fig:dif_ang}
\end{figure}

{\it Theoretical rate coefficients.}
Figure~\ref{fig:dif_ang} shows the rate coefficient $k_3(E_c,\alpha)$ as functions of collision energy obtained for different values of angle $\alpha$ between O$_2$ and the direction to Ar. 
Comparing the upper set of data (indicated by circles), which is associated with the ozone molecules of all binding energies, and the lower set (squares), i.e.,  the formation rates of ``deeply-bound'' levels of O$_3$ with binding energies larger than 200~cm$^{-1}$~($\approx$ 288~K), one can conclude that the majority of molecules formed through ternary recombination are highly-excited. Such molecules can be destroyed or stabilized in collisions with other species present in the gas. The cross sections for the two processes are comparable. The radiative stabilization is negligible because of the very long radiative lifetimes of these states. Therefore, comparing with an experiment at 300~K, in which only deeply-bound or stabilized O$_3$ molecules are accounted for, one should consider the lower set of data.

\begin{figure}
	\centering
	\includegraphics[scale=0.41]{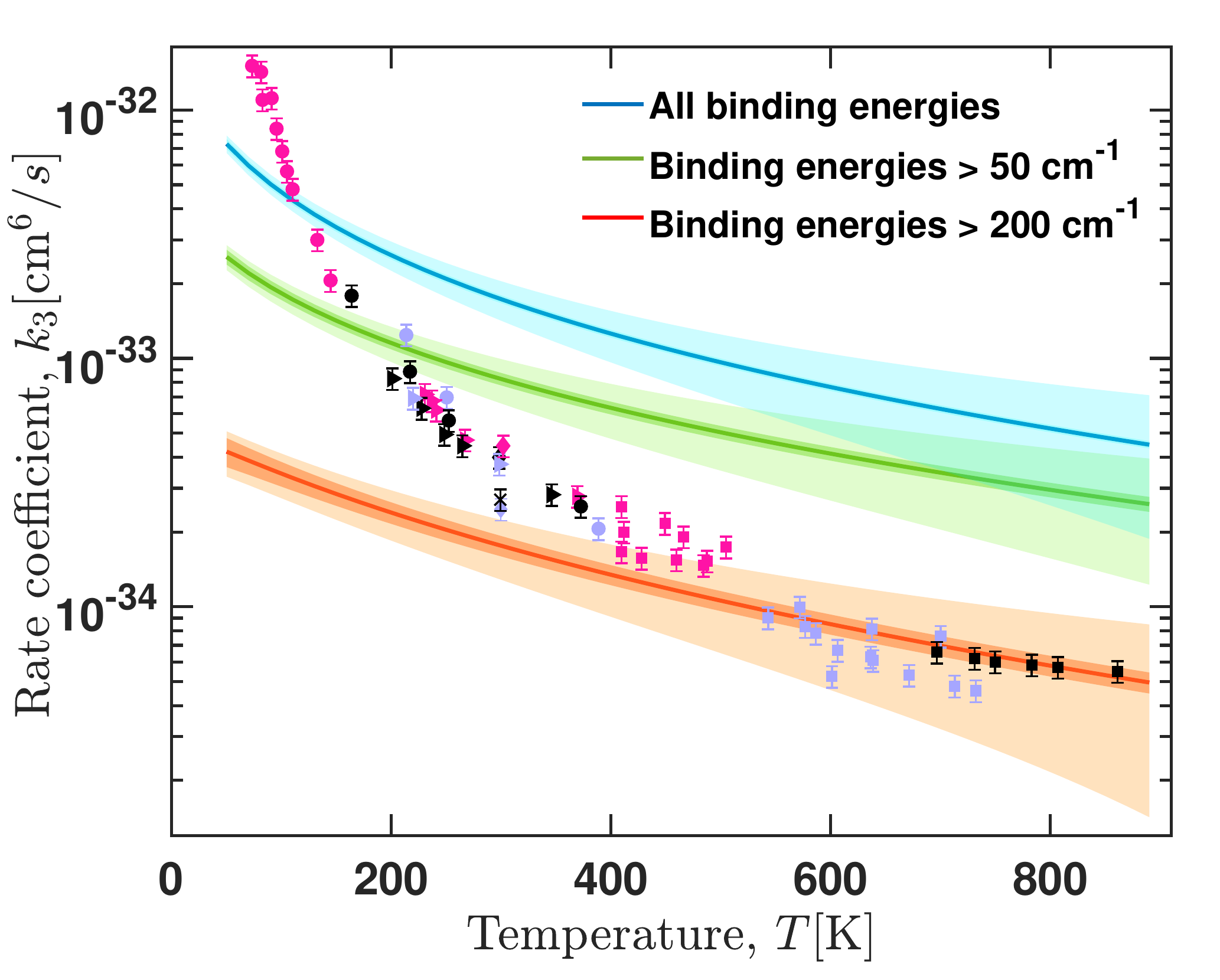}
	\caption{Thermally-averaged recombination rate coefficients for formation of molecules with all binding energies (blue data), with binding energies $>50$ cm$^{-1}$ (green data), and with $>$200 cm$^{-1}$ (red data). The dark shaded area indicates the statistical error due the stochastic nature of the Monte Carlo technique used in the calculations. The light shaded area shows the confidence interval due to both statistical error and the effect of neglecting explicitly $\alpha$. Experimental data for $k_3(T)$: magenta circles -- Ref.~\cite{rawlins1987dynamics}, black circles -- \cite{hippler1990temperature}, cornflower blue circles -- \cite{mulcahy1968kinetics} (scaled to the value at 300~K from Ref.~\cite{kaufman1967m}), black triangles -- \cite{huie1972absolute}, cornflower blue triangles -- \cite{arnold1979temperature}, magenta squares -- \cite{Intezarova1967} (as cited in Ref.~\cite{LUT05:2764}), cornflower blue squares -- \cite{park1977reaction}, black squares -- \cite{sauer1965pulse}, cornflower blue diamond -- \cite{center1975shock} (as cited in Ref.~\cite{LUT05:2764}), black $\times$ -- \cite{bevan1973kinetics}, black diamond -- \cite{kaufman1967m}, magenta diamond -- \cite{slanger1970reaction}, magenta triangles -- \cite{klais1980reinvestigation}.}
	\label{fig:k3T_err_CT}
\end{figure}

Figure~\ref{fig:k3T_err_CT} gives thermally-averaged  rate coefficients $k_3(T)$ for molecules with all binding energies and for those with binding energies larger than 50~cm$^{-1}$~($\approx$~72~K) and 200~cm$^{-1}$~($\approx$~288~K) and compares them with available experimental data. Clearly, the theoretical coefficients have different $T$-dependence due to the fact that in an experiment at a given temperature $T$, only the molecules with binding energies larger than $T$ are collisionally stabilized and accounted for. Indeed, the theoretical data for molecules with binding energies $>$50~cm$^{-1}$ and $>$200~cm$^{-1}$  agree with the experimental data at 72~K and 288~K within about 50\%. 


\section{\label{sec:disc}Discussion}

\begin{figure}
	\centering
	\includegraphics[scale=0.41]{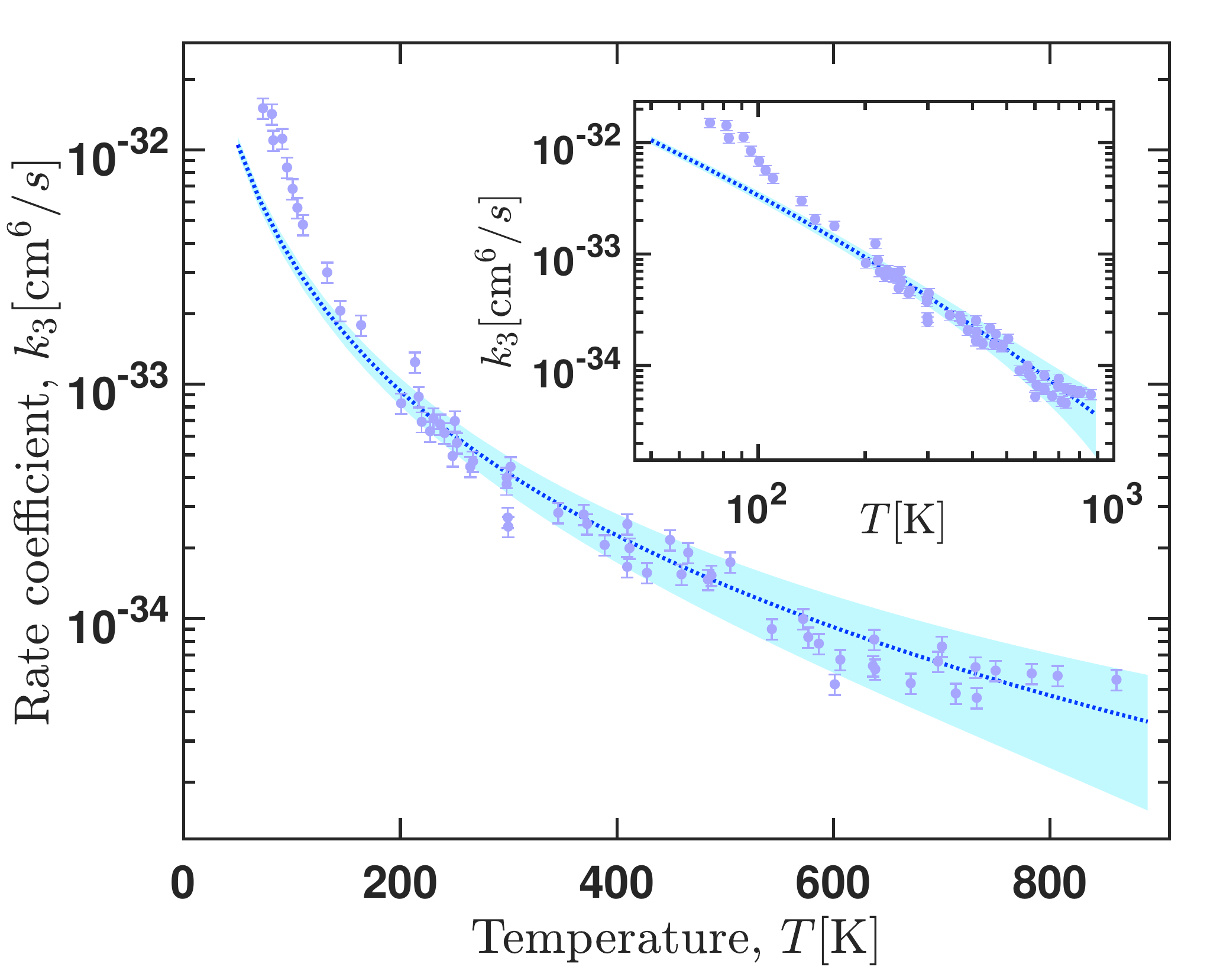}
	\caption{Thermally-averaged recombination rate coefficients for formation of molecules with all binding energies scaled by the factor of 50~cm$^{-1}/T$ accounting for the process of vibrational quenching. The inset  shows the same data in the log-log scale.}
	\label{fig:k3T_QU}
\end{figure}

 To compare the obtained theoretical results with the experimental data more accurately, one would have to account for processes taking place in the experiment once the ternary recombination forms an initial distribution over vibrational levels of O$_3$, the nascent population. The leading process is the vibrational de-excitation of highly-excited O$_3$ molecules in collisions with other species (vibrational quenching). A detailed study of vibrational quenching is underway but beyond the scope of this work. Here, we will account for the process using a statistical factor $\Delta E/T$, suggested \cite{LUT05:2764,troe1977theory} for the stabilization process, which is the same as the vibrational quenching of highly-excited O$_3$. The averaged value $\Delta E$ of energy transfer during the stabilization is of the order of the energy of  Ar-O$_3$  interaction averaged over involved vibrational states of O$_3$, which is about 50~K (see the Ar-O and Ar-O$_2$ curves in Fig.~\ref{fig:pots}). Therefore, we use $\Delta E=50$~K. The resulting rate coefficient, accounting for the vibrational quenching, is shown in Fig.~\ref{fig:k3T_QU}. As one can see, the agreement with the experimental data is remarkable for temperatures above 70~K. The good agreement indicates that despite several approximations made in the model, the main physics of the process is accounted for correctly. 

%
%
%

\begin{figure}
	\centering
	\includegraphics[scale=0.42]{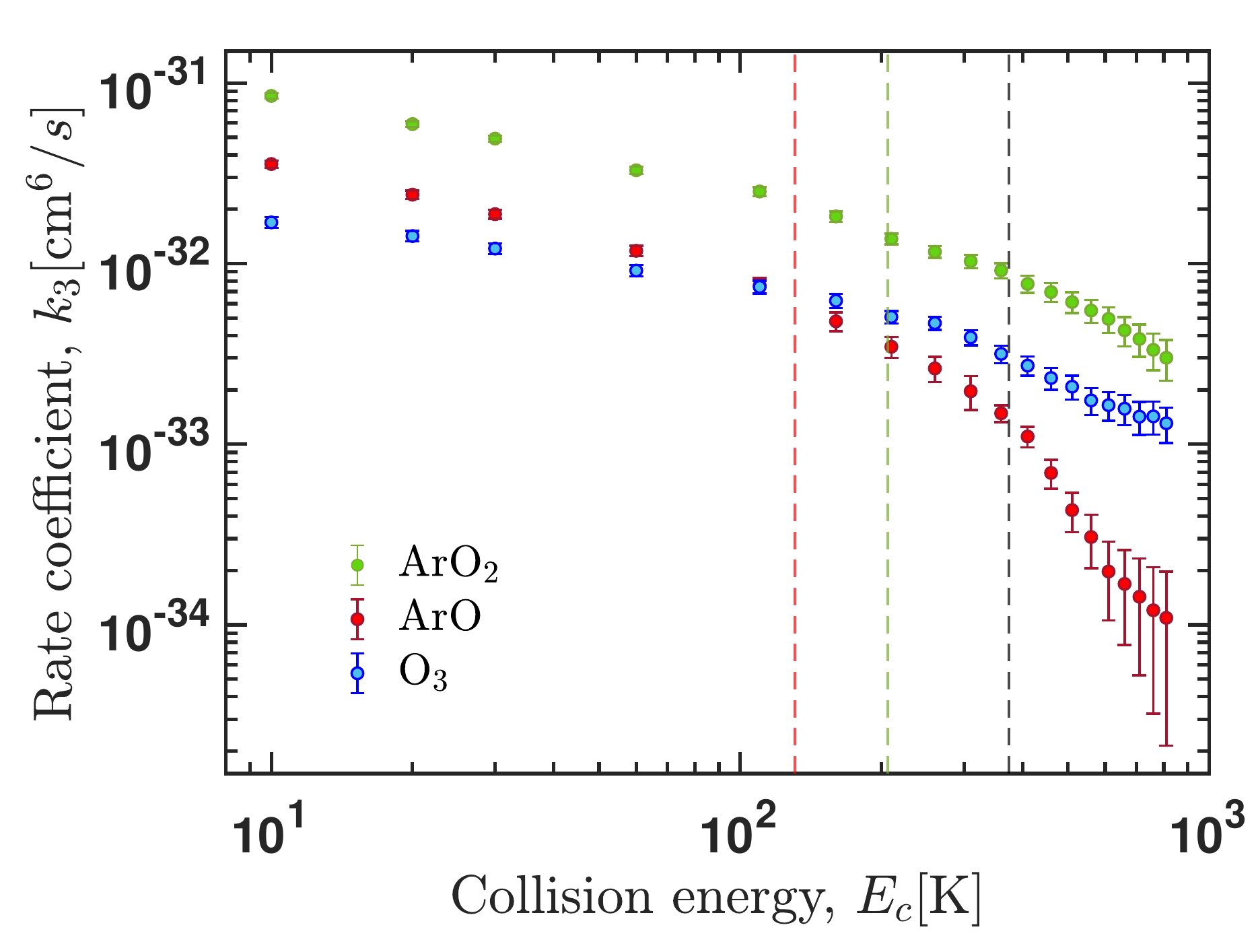}
	\caption{Ternary recombination rate coefficient $k_3(E_c)$ for formation of the three possible products in the O+O$_2$+Ar collisions: O$_3$, ArO$_2$, and ArO. Red and green dashed lines indicate the dissociation energies of ArO and ArO$_2$, respectively. The black dashed line shows the energy related to shoulder structure of O-O$_2$.}
	\label{fig:k3T_allP}
    \end{figure}

{\it Two-step models: stabilization and Chaperon mechanisms.} The present results allow us to analyze which of the two-step models (stabilization or Chaperon) is dominant. Figure~\ref{fig:k3T_allP} gives the rate coefficients for the formation of all three possible products of the reaction, O$_3$, ArO$_2$, and ArO. The figure demonstrates two distinct characteristics: First, at low collision energies ($E_c\lesssim 130$~K) \cite{Mirahmadi2021, Mirahmadi2021a},  ArO$_2$ is formed more quickly than ArO, which is formed more quickly than O$_3$ (assuming equal densities of Ar, O, and O$_2$). The result is explained by the fact that the attractive potential of the Ar-O$_2$ system decays slower than that of Ar-O, so that it has a larger density of available vibrational states to be formed (a larger phase-space volume in the classical formulation of the problem). The same argument applies to the comparison between the O-O$_2$ and Ar-O potentials. 

The second distinct feature in Fig.~\ref{fig:k3T_allP} is the change in the power-law behavior of recombination rates at 200-300~K. The first sudden drop appears in $k_3$ of ArO around its dissociation energy: At collision energies larger than the dissociation energy, fewer ArO molecules are formed due to higher initial velocities. The rates of formation of ozone and ArO$_2$ are also affected in a similar way, although in these cases, slopes of the coefficients vary much less than that for ArO. There is another drop in the recombination rates of all products when the collision energy surpasses the dissociation energy of ArO$_2$. Interestingly, there is another smoother change in the slope of the rate coefficient for O$_3$ formation that may be due to the shoulder structure on O$_3$ \cite{TYU13:134307,lapierre16}, visible in Fig.~\ref{fig:pots}. The overall effect is that for collision energies $\gtrsim150$ K, the probability of forming bound levels of ArO decreases significantly (compared with other products), and the probability of ozone formation overcomes it. 

The more significant rate coefficients for the formation of  ArO$_2$ and ArO compared to O$_3$ at low collision energies mean that the Chaperon mechanism is more important than the stabilization at temperatures $\lesssim$ 100~K. At higher temperatures, the probability of forming weakly bound ozone molecules is more prominent than for ArO, and hence the stabilization mechanism starts to play a role. Similarly, at temperatures above 1000~K, typical collision energies are larger than binding energies of ArO$_2$ and ArO, so that formation of the corresponding bound molecules is less likely, and the stabilization mechanism becomes dominant. 

\section{Summary and conclusions}
We would like to stress the following findings of the present study.
\begin{enumerate}
 \item 
The rate coefficients for the formation of O$_3$ in ternary collisions O+O$_2$+Ar, without using any two-steps approximation, were computed for the first time as a function of collision energies. Thermally-averaged coefficients were derived for temperatures 5-900~K. 
 \item 
Due to a relatively long observation time in experiments that were performed so far, a direct measurement of the ternary rate coefficient was not possible since the majority of O$_3$ molecules formed initially, the nascent population, are weakly-bound and, therefore, destroyed or stabilized in collisions with other species present in the gas (Ar, O$_2$, etc.) that affected the observation results.
 \item 
Accounting for the process of vibrational quenching of the nascent population using the statistical formula by Troe \cite{troe1977theory}, an excellent agreement with available experimental data for temperatures 100-900~K is obtained. At temperatures 50-100~K theory gives smaller rate coefficients. It is likely that the classical-trajectory approach is not applicable for these temperatures because at the corresponding collision energies, details of the resonant structure of O$_3$ and, possibly, ArO$_2$ metastable rovibrational states could be important, which cannot be treated accurately using  classical-trajectory methods.
 \item 
The rate coefficients obtained in an explicitly three-body approach, without assuming any two-step mechanism, allowed us to conclude that the Chaperon mechanism in ozone formalism is dominant at temperatures below 100~K, while the energy-transfer stabilization mechanism appears to be dominant above 1000~K.
 \end{enumerate}
 
\section{Methods}
{ArO$_3$ \it PES.}
The Ar-O$_2$ and Ar-O intermolecular PESs were constructed using the unrestricted version of the explicitly-correlated single- and double-excitation coupled cluster method with a perturbative treatment of triple excitations [UCCSD(T)-F12a] with the augmented correlation-consistent triple-zeta {\texttt aug-cc-pVTZ} basis set. The final root mean square error of the fit of the {\it ab initio} data was within 1 cm$^{-1}$ at all energies in the PES potential wells. 


{\it Dynamics.}
Reduced to three-body problem, the system is described by six-dimensional (6D) hyperspherical coordinates. The initial angular dependence, which depends on both direction and magnitude of impact parameter $\vec{b}$ and initial momentum $\vec{P}_0$ vectors, has been averaged out by means of the Monte Carlo method. The initial hyperangles determining the orientation of these vectors in the 6D space are sampled randomly from probability distribution functions associated with the appropriate angular elements in hyperspherical coordinates (see, e.g., Refs.~\cite{Perez-Rios2014,Perez-Rios2020}). For the results reported in this work, $1.3 \times 10^9$ random initial configurations of ($\vec{b}$, $\vec{P}_0$) pairs has been sampled.

To solve the Hamilton's equations we made use of the variable-step, variable-order Adams-Bashforth-Moulton method of order 1 to 13, which is based on a multi-step predictor–corrector algorithm \cite{Shampine1997,Ashino2000,Shampine2002}. 
The acceptable error for each time-step has been determined by absolute and relative tolerances equal to $10^{-15}$ and $10^{-11}$, respectively. The total energy is conserved during collisions to, at least, three significant digits and the magnitude of total angular momentum vector is conserved to at least six significant digits. The initial magnitude of hyper-radius is generated randomly from the interval $[R_0-20a_0, R_0+20a_0]$ centered around $R_0 = 200 ~ a_0$. This value fulfills the condition for three particles to be initially in an uniform rectilinear state of motion.

{\it Evaluation of the confidence interval.}
To calculate the confidence interval we considered both the statistical error, owing the inherent stochastic nature of the Monte Carlo technique, and the error due to different $\alpha$ angles. The latter one has been implemented as the standard deviation of the results obtained by using $\alpha=\pi/2$, from the weighted average over different angles, i.e., 

\begin{equation}
\left\langle k_3(E_c)\right\rangle_\alpha = \frac{1}{2} \int_{0}^{\pi}k_3(E_c,\alpha)\sin\alpha d\alpha .    
\end{equation}

\noindent
Here $E_c = P_0^2/(2\mu)$ is the collision energy with $\mu$ being the three-body reduced mass. Figure~\ref{fig:dif_ang} shows rate coefficients $k_3(E_c,\alpha)$ as functions of collision energy obtained for different values of angle $\alpha$ between O$_2$ and the direction to Ar.

\section{Acknowledgements}
This work acknowledges the support from the Russian Science Foundation grant \textnumero 19-12-00171 and the National Science Foundation, Grant No.2110279.

%

\bibliographystyle{apsrev4-2}
\bibliography{O3.bib}

\end{document}